\documentclass[useAMS,usenatbib]{mn2e}
\usepackage[dvips]{graphicx}
\input{epsf}

\title[Extragalactic radio sources with upturn spectrum]{Multi--frequency
spectral analysis of extragalactic radio sources in the 33--GHz VSA catalogue:
sources with flattening and upturn spectrum.}

\author[Tucci et al.]{M. Tucci$^1$, J.A. Rubi\~{n}o-Martin$^1$, R. Rebolo$^1$, 
R. Genova-Santos$^2$, R.A. Watson$^3$, 
\newauthor R.A. Battye$^3$, K.A. Cleary$^4$, R.D. Davies$^3$,
R.J. Davis$^3$, K. Grainge$^2$, M. Hobson$^2$,
\newauthor R.D.E. Saunders$^2$, A. Scaife$^2$, P.F. Scott$^2$ \\
$^1$Instituto de Astrofis\'{\i}ca de Canarias, 38200 La Laguna, Tenerife,
Canary Islands, Spain \\
$^2$Astrophysics Group, Cavendish Laboratory, University of Cambridge,
Madingley Road, Cambridge CB3 0HE \\
$^3$Jodrell Bank Observatory, University of Manchester, Macclesfield, Cheshire
SK11 9DL \\
$^4$California Institute of Technology, 1200 E California Blvd, Pasadena,  
CA 91125}

\begin{document}

\date{}

\pagerange{\pageref{firstpage}--\pageref{lastpage}} \pubyear{2007}

\maketitle

\label{firstpage}

\begin{abstract}

We present an analysis of the spectral properties of the extragalactic
radio sources in the nearly--complete VSA sample at 33\,GHz. Data from
different surveys are used to study source spectra between 1.4 and 33\,GHz.
We find that, in general, spectra can not be well described by a single power
law in the range of frequencies considered. In particular, most of the VSA
sources that are steep between 1.4 and 5\,GHz, show a spectral flattening at
$\nu>5\,$GHz. We identify 20 objects (19\% of the sample)
clearly characterized by an upturn spectrum, i.e., a spectrum falling
at low frequencies ($\nu\la5\,$GHz) and inverted at higher
frequencies. Spectra with high--frequency flattening or upturn shape
are supposed to occur when the emission from the AGN compact core begins
to dominate over the component from extended lobes. This picture fits
well with the AGN unified scheme, for objects observed at intermediate
viewing angles of the AGN jet. Finally, we discuss implications that this class
of sources can have on future CMB observations at high resolution.

\end{abstract}

\begin{keywords}

Galaxies: active -- quasars: general; Radio continuum: galaxies; Cosmic
microwave background

\end{keywords}

\section{Introduction}
In recent years, extragalactic radio sources have received increased
attention due to their contamination of the cosmic background radiation
(CMB) temperature fluctuations and polarization on small angular scales 
(\citealt{tof05}; \citealt{tuc04}). At microwave wavelengths, this contribution
is dominated by objects powered by active galactic nuclei (AGNs), such as
quasars or BL Lacs. These sources are usually separated according to the
slope of the spectrum at GHz frequencies (usually assumed a power law;
hereafter $S\propto\nu^{\alpha}$): steep--spectrum if the spectral index
$\alpha\le-0.5$ and flat--spectrum otherwise. The sources are thus divided into
two classes, corresponding approximately to those where the radio flux is
predominantly emitted from the extended lobes and those where the compact core
is the brightest component \citep{pea85}. 

Based on this scheme, evolutionary models for radio sources provide accurate
predictions for the number counts at frequencies $\nu\la8\,$GHz and down to
flux densities of fractions of mJy (\citealt{dun90}; \citealt{tof98};
\citealt{jac99}; \citealt{dez05}). Recently, \citet{dez05} updated and improved
evolutionary models taking into accounts new high--frequency data from the
15--GHz 9C \citep{wal03}, 18--GHz ATCA catalogues \citep{ric04} and WMAP (first
year; \citealt{ben03}). However, estimates of the radio source contribution to
temperature fluctuations at cm/mm wavelengths, i.e. at wavelengths interesting
for CMB observations, is still problematic. In fact, radio spectra of AGNs can
be much more complex than a single power law, that may not provide a
satisfactory description in a wide range of frequencies (see, e.g.,
simultaneous multi--frequency measurements of radio spectra by \citealt{tru03}
and \citealt{bol04}). Different mechanisms can be responsible for this: (i) a
spectral steepening is expected due to the more rapid energy loss of
high--energy electrons with source age; (ii) a transition from optically thick
to optically thin regimes can
occur at high frequencies; (iii) at different wavelengths radio emission can be
dominated by different components characterised by a distinct spectral
behaviour. For example, AGNs with spectra peaked at GHz frequencies
(GPS) are well known, while objects with spectra still inverted at frequencies
as high as 20--30\,GHz are expected to fall at some peak frequency (see
\citealt{ode98} and \citealt{dez05} for reviews on GPS and extreme GPS
sources).

Recently, indications of a class of radio sources with spectra that are
``opposite'' in respect to the GPS ones, are appearing in the literature.
\citet{sad06}, using multi--frequency data of the ATCA 20--GHz pilot survey,
observed objects with spectrum falling at low frequencies, but turning up and
begining to rise around 5\,GHz, and called them ``upturn'' sources. They found
that about 20\% of the sources belongs to this class. \citet{tru03} identified
in the first--year WMAP source catalogue 15 sources (corresponding to 7\%) with
``combined spectra with a power low--frequency component and a flat (inverse)
high--frequency component''. Upturn objects are observed also in recent
observations by ATCA (\citealt{mas07}; \citealt{sad07}) and by \citet{par07} in
nearby galaxy clusters.

The presence of a significant fraction of
upturn sources in samples selected at few tens of GHz could have important
implications for CMB observations and for models predicting the radio source
contribution at CMB frequencies. These sources, in fact, could appear in
high--frequency surveys with flux densities much brighter than expected from
low--frequency extrapolations.

In this paper we carry out a multi--frequency spectral analysis of the sources
detected by the extended version of the Very Small Array (VSA) at 33\,GHz
(\citealt{cle05}, C05 hereafter). VSA provides a deep and nearly--complete
survey of extragalactic radio sources with a
flux limit of 20\,mJy. A first discussion on the spectrum of these sources and
their variability was carried out by C05, with an estimate of the source
number counts at 33\,GHz in the flux range 20--114\,mJy. Here, we exploit
surveys at different frequencies, as the NRAO VLA Sky Survey (NVSS,
\citealt{con98}) at 1.4\,GHz, the Green Bank survey (GB6, \citealt{gre96}) at
4.8\,GHz and the Ryle Telescope (RT, \citealt{wal03}) at 15\,GHz, in
order to study
the spectrum of VSA radio sources in a quite extended range of frequencies.
In particular, we focus our attention on sources whose spectrum shows
a high--frequency flattening or an upturn behaviour, discussing a
physical interpretation for such spectra.

\section{The VSA source catalogue}

The Very Small Array (VSA) consists of a 14--element interferometer operating
in the band 26--36\,GHz, located at Teide Observatory in Tenerife (see,
e.g., \citealt{wat03}). Two 3.7\,m antennas, spaced 9.2\,m apart (called the
``source subtractor''), are used to measure the flux density of sources present
in the VSA fields. The sources to be followed by the subtractor are preselected
from a catalogue at 15\,GHz provided by the Ryle Telescope (RT, see e.g.
\citealt{jon91} and \citealt{wal03}) that surveyed the VSA fields in advance of
CMB observations. For the extended configuration of the VSA (\citealt{gra03};
\citealt{dic04}) it was required that all sources with flux $S\ga20\,$mJy were
subtracted from CMB maps produced by the VSA main array. In order to satisfy
this goal, the source subtractor followed all the sources detected by RT in the
VSA fields up to the RT flux limit of 2\,mJy. This ensures that all sources
(except for extremely inverted ones) with $S\ge20\,$mJy at 33\,GHz are observed
by the source subtractor (C05).

Our analysis is based on the catalogue of extragalactic radio sources
presented by C05 and used to estimate the number counts at 33\,GHz and
to correct the source contribution to the VSA extended array power
spectrum (see also \citealt{dic04}). Due to errors in the text of Table A1 in
C05, we use the list of 33--GHz sources from \citet{cle08}.
The catalogue consists of 102 
sources with flux density between 20 and 114\,mJy, within well--defined
regions of the VSA fields that cover a total area of 0.044\,sr. We
consider also 3 further objects present in the VSA area and brigther
than 114\,mJy. All these
sources satisfy the pre--selection condition to be brighter than
10\,mJy at 15\,GHz. As a consequence, the sample can not be considered
strictly complete because there is the possibility that some sources
with rising spectral index between 15 and 33\,GHz
($\alpha_{15}^{33}\ga1$) and $S_{33}\ge20\,$mJy exist in the VSA
fields but it is not present in the sample. However, such inverted
sources are rare and the loss of few of them would not affect the
following analysis.

\begin{figure}
\includegraphics[width=90mm]{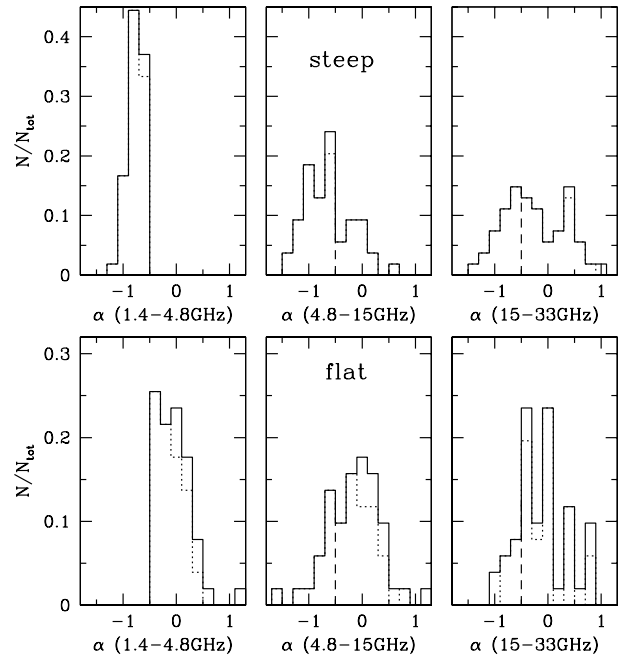}
\caption{Distribution of spectral indeces for steep-- and flat--spectrum
(defined between [1.4,\,4.8]\,GHz) sources, calculated in frequency ranges
[1.4,\,4.8]\,GHz, [4.8,\,15]\,GHz and [15,\,33]\,GHz. Dotted histograms are
excluding sources without 4.8--GHz measurements. Vertical dashed lines
correspond to $\alpha=-0.5$.}
\label{f1}
\end{figure}

\section{Spectral properties in the frequency range [1.4,\,33]\,GHz}

\begin{figure*}
\includegraphics[width=85mm]{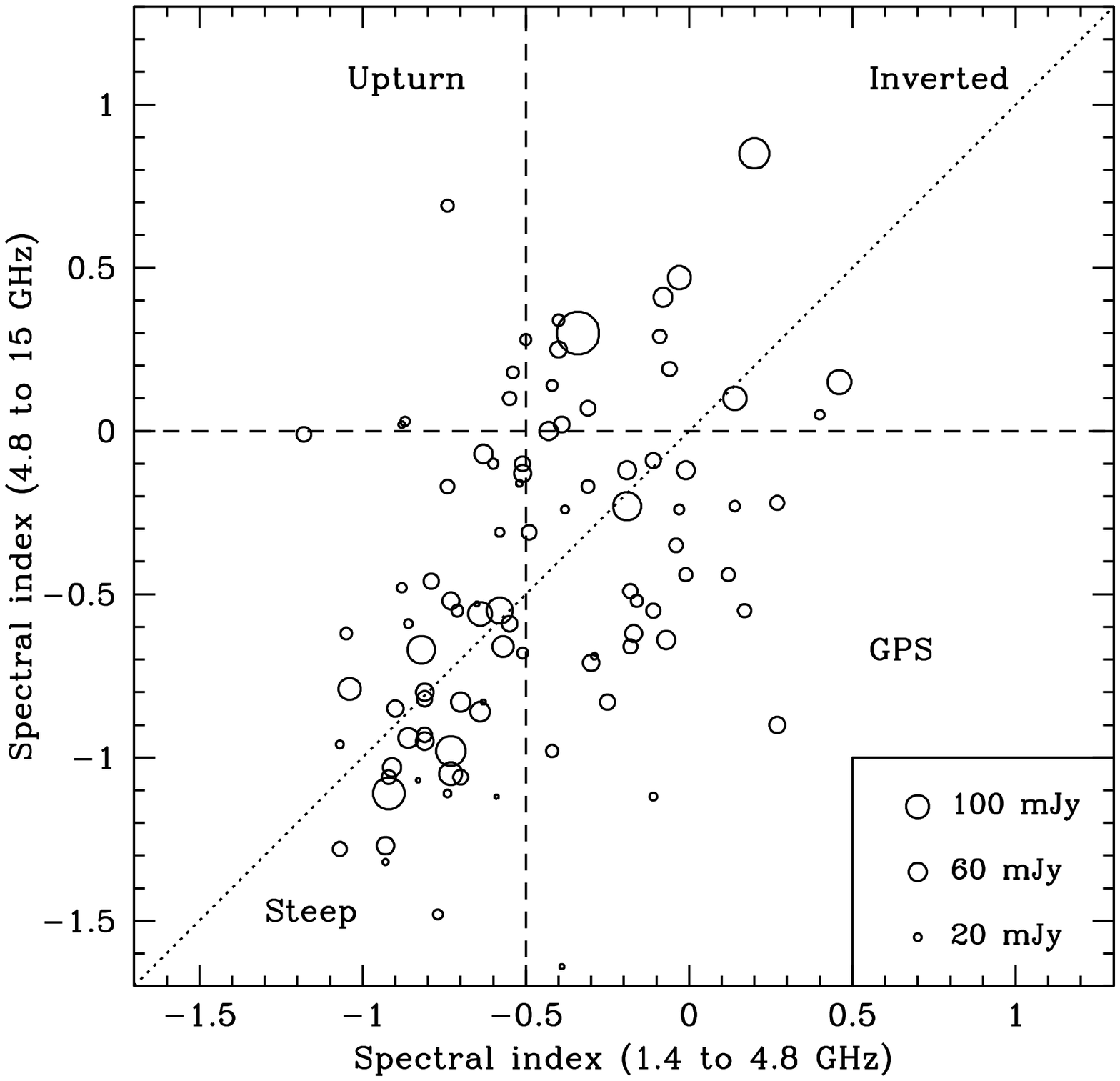}
\includegraphics[width=85mm]{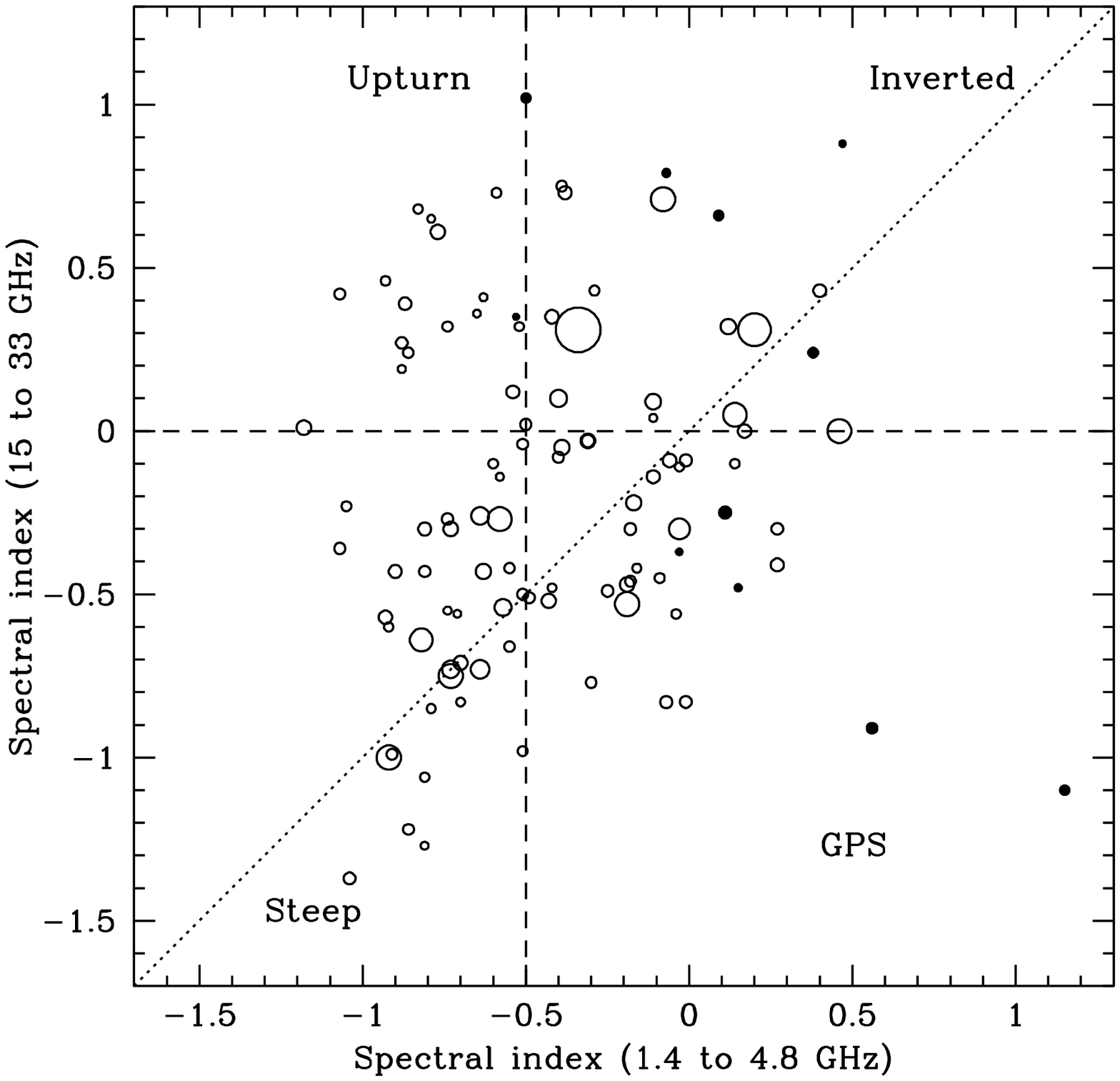}
\caption{``Two colour diagrams'': the spectral index calculated at intermediate
(left plot) and high (right plot) frequencies vs the value at low frequencies. 
The size of the circle is related to the flux densitiy at 15\,GHz (left plot)
and at 33\,GHz (right plot). The vertical dashed line divides sources into
steep-- and flat--spectrum. Full dots in the right plot correspond to sources
without 4.8\,GHz flux; in this case, $\alpha_{1.4}^{4.8}$ is substituted with
$\alpha_{1.4}^{15}$.}
\label{f2}
\end{figure*}

Spectral properties of sources in the VSA sample can be studied using flux
densities from surveys at lower frequencies: at 15\,GHz from the RT; at 
4.8\,GHz from the GB6; and at 1.4\,GHz from the NVSS. Only 11 objects do not
have a counterpart in the GB6 survey, most of them characterized by an inverted
spectrum between 1.4 and 15\,GHz and a low flux density at 1.4\,GHz.
We separate VSA sources into steep-- and flat--spectrum sources according
to their spectral index between 1.4 and 4.8\,GHz (or 15\,GHz in the case that
4.8--GHz flux density is not available). The sample is almost equally divided
into steep-- and flat--spectrum sources, being 54 and 51 respectively, as
predicted by evolutionary models for radio sources in this flux--density range
(see, e.g., \citealt{dez05}).

Figure \ref{f1} shows the distribution of spectral indices for the two source
populations in three different intervals of frequencies: $[1.4,\,4.8]$\,GHz,
$[4.8,\,15]$\,GHz and $[15,\,33]$\,GHz (hereafter indicated as low,
intermediate and high frequencies respectively). We suppose
$\alpha_{1.4}^{4.8}=\alpha_{4.8}^{15}=\alpha_{1.4}^{15}$ for the 11 sources
without measurements at 5\,GHz. For {\bf flat--spectrum sources}, the 
dispersion in the distribution of spectral indices increases at intermediate
and high frequencies, with tails up to strongly inverted or steep indices
($|\alpha|\ga1$). However, the average value of $\alpha$ remains nearly
constant with the frequency interval,
$$
\langle\alpha_{1.4}^{4.8}\rangle=-0.05~~~~
\langle\alpha_{4.8}^{15}\rangle=-0.15~~~~
\langle\alpha_{15}^{33}\rangle=-0.11
$$
Different remarks can be drawn for {\bf steep--spectrum sources}. The
distribution of $\alpha_{1.4}^{4.8}$ is very peaked between -1 and -0.5, as
expected because very steep--spectrum sources have low probability to be
observed at high frequency. In the intermediate interval spectral indices 
spread both to positive and to very steep values, but keeping
a similar average $\alpha$. On the contrary, a clear ``flattening'' in spectra
is observed at high frequencies, where more than 60\% of sources have
$\alpha_{15}^{33}>-0.5$ and 1/3 of steep sources has become inverted; the
average value increases by $+0.5$ from the low-- to the high--frequency
interval:
$$
\langle\alpha_{1.4}^{4.8}\rangle=-0.75~~~~
\langle\alpha_{4.8}^{15}\rangle=-0.65~~~~
\langle\alpha_{15}^{33}\rangle=-0.25
$$

The presence of curvatures in source spectra is more evident in
``two--colour diagrams'': in Figure \ref{f2} we compare the
low--frequency spectral index, $\alpha_{1.4}^{4.8}$, with the one
obtained at higher frequencies. In the left plot of Figure \ref{f2},
we have $\alpha_{4.8}^{15}$ vs $\alpha_{1.4}^{4.8}$: spectral indices
seem to be distributed around the line $\alpha_{4.8}^{15}=\alpha_{1.4}^{4.8}$,
with a bigger dispersion for flat--spectrum sources. Only in a few objects does
the spectral index significantly change. No
systematic trends can be observed and a single power law could statistically
describe source spectra. This is not the case when we consider the
high--frequency spectral index ($\alpha_{15}^{33}$ vs $\alpha_{1.4}^{4.8}$; see
right plot). In most of the steep--spectrum sources the spectrum tends to
flatten at high frequency or even to become inverted (i.e. with the typical
upturn shape). For flat--spectrum sources we notice a clear flattening only in
objects with spectral index $-0.5<\alpha_{1.4}^{4.8}<-0.3$, that represent a
transition from proper steep--spectrum (lobe--dominated AGN) and flat--spectrum
(core--dominated AGN) sources. In general, if we
consider sources with $\alpha_{1.4}^{4.8}\ga-0.3$, there is an average
steepening in spectra at $\nu>5$\,GHz and very few objects keep the spectrum
flat or inverted at high frequencies.

The radius of circles in Figure \ref{f2} is related to the flux density of
sources: a visual inspection of the plots does not show any correlation between
the position of sources in the two--colour plot and their flux density. To
confirm this, we calculate the {\it linear correlation coefficient} $r$ between
the flux density at 33\,GHz, $S_{33}$, and the spectral index variation
$\alpha_{15}^{33}-\alpha_{1.4}^{4.8}$: using all the sample, we find $r=0.027$
with a propability for the null hypothesis of zero correlation of 0.76. This
probability increases up to 0.8 (0.96) if we consider only sources with
$\alpha_{1.4}^{4.8}<0$ ($<-0.3$). This is particularly important because the
lack of correlation rules out that spectral flattening at high frequencies
occurs only in faint objects, where uncertainties in flux densities are higher:
as example, we find that the brightest source of the sample (source 1734+3857)
have $\alpha_{15}^{33}\ga\alpha_{4.8}^{15}\gg\alpha_{1.4}^{4.8}$. Moreover, a
correlation between 33--GHz flux densities and spectral flattening should be
expected if flattening is observed in objects that are just in the extreme of
the spectral index distribution.

\begin{table}
\caption{Number of sources with $|\Delta\alpha|=|\alpha_H-\alpha_L|>0.3$ (in
brackets, $>0.5$).}
\begin{tabular}{@{}ccccccc@{}}
\hline
 & \multicolumn{6}{c}{$\alpha_H$ vs $\alpha_L$} \\
\hline
 $\alpha_L$ & N & steeper & flatter & upturn & GPS & $\langle\alpha\rangle$ \\
\hline
 $\le-0.5$ & 53 & 3 (0) & 35 (25) & 19 & -- & 0.55 \\
 $(-0.5,\,-0.3]$ & 16 & 4 (3) & 8 (6) & 6 & -- & 0.16 \\
 $>-0.3$ & 36 & 21 (11) & 5 (3) & 2 & 10 & -0.33 \\
\hline
\end{tabular}
\label{t0}
\end{table}

\subsection{Fit by two power law of source spectra}

\begin{figure}
\includegraphics[width=80mm]{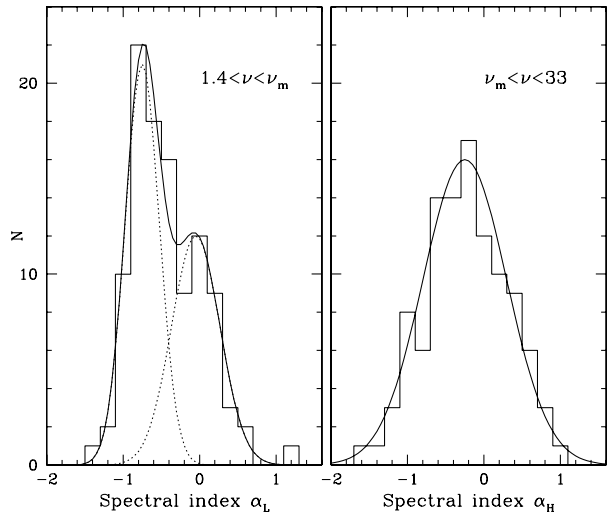}
\caption{Distribution of spectral indices at low (left plot) and high (right
plot) frequencies. Solid lines are the Gaussian fit of the distributions.}
\label{f300}
\end{figure}

\begin{figure}
\includegraphics[width=90mm]{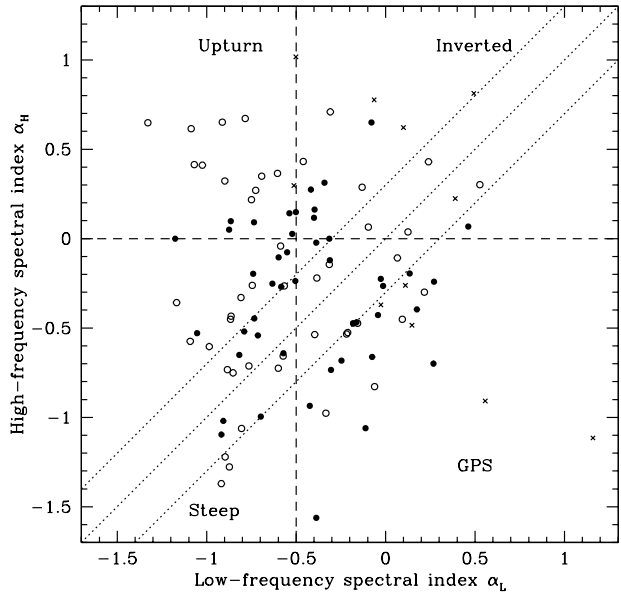}
\caption{Two colour diagram using spectral indices from two--power--law best
fit. Open points are for objects whose $\alpha_H$ is calculated between 15 and
33\,GHz; solid points for objects whose $\alpha_H$ is calculated between 4.8
and 33\,GHz; crosses for objects without 4.8--GHz measurements. The diagonal
dotted lines represent $\alpha_H=\alpha_L+\,$0.3/0/--0.3 respectively.}
\label{f3}
\end{figure}

For a better characterization of source spectra and their curvature (when
present), we suppose spectra to be well described by two different
power laws. Let $\nu_m$ be the frequency at which the change in the spectral
slope occurs. We define the low--frequency ($\alpha_L$) and the high--frequency
($\alpha_H$) spectral indices as the indices of the two best--fit power laws
between [1.4,\,$\nu_m$]\,GHz and [$\nu_m$,\,33]\,GHz respectively. The first
step is to find out the frequency $\nu_m$ at which source spectra are better
fitted, considering that, in our case, 
$\nu_m$ can only be 4.8 or 15\,GHz. Figure \ref{f300} shows the distribution
of $\alpha_L$ and $\alpha_H$ of the VSA sources. The former presents a
bimodal distribution that can be fitted by two Gaussian distributions of 
mean $-0.75$ and $-0.05$ and dispersion 0.22 and 0.32 respectively.
The two components can represent the two populations of AGN, steep-- and
flat--spectrum sources, clearly separated at low frequencies. On the
contrary, at high frequencies the spectral index distribution is well
fitted by only one Gaussian distribution with mean $-0.25$ and dispersion
0.55, as result of the average flattening of steep--spectrum sources and the
average steepening of flat--spectrum sources at high frequencies.

The spectral properties of VSA sources between 1.4 and 33\,GHz can be now
described by a single ``two--colour diagram'' using $\alpha_L$ and $\alpha_H$
(see Figure \ref{f3}). There we plot open (full) points for sources with
$\nu_m=15$ (4.8)\,GHz; sources without 4.8--GHz measurements are
indicated by a cross. Uncertainties in the spectral index are usually
around 0.1--0.3 for both $\alpha_L$ and $\alpha_H$; however, they can
increase more than 0.3 when $\alpha_H$ is calculated only by 15 and
33--GHz data (open points), due to the uncertainties associated with the VSA
measurements.

The majority of sources (about 3/4; see Table \ref{t0}) are observed to change
their spectral slope more than 0.3, indicating a possible curvature in the
spectrum, while very few sources can be fitted by a single power law. An 
average flattening in steep--spectrum sources ($\alpha_L\le-0.5$) is still
confirmed:
66\% of these sources have $\Delta\alpha=\alpha_H-\alpha_L\ge0.3$, and nearly
50\% have $\Delta\alpha\ge0.5$. From Figure \ref{f3} we notice that the largest
changes in $\alpha$ occur for sources with $\nu_m=15\,$GHz (i.e. open points or
crosses): these large differences could be overestimated due to the
uncertainties associated with the VSA measurements and some caution is
required. In the next section, we discuss about upturn sources and we deal with
this issue.

It is important to stress that few objects that are flat spectrum at GHz
frequencies, keep their spectral slope flat or inverted up to 33\,GHz in Figure
\ref{f3}. Instead, a general steepening is found. Looking at sources with
$\alpha_L\sim0$, the typical steepening is about
$\Delta\alpha\simeq0.3$--0.5. This result seems to indicate that, contrary to
what normally assumed, a single power law with index $\alpha\approx0$ is not,
even statistically, a correct description of spectra for flat--spectrum sources
at high frequencies. The contribution of flat-spectrum sources at the
frequencies used for CMB observations could thus be lower than predicted. This
is in agreement with results from \citet{wal07}, with recent observations by
the ATCA telescope (\citealt{mas07}; \citealt{sad07}) and with spectral
properties of WMAP sources \citep{gon08}.

\begin{figure}
\includegraphics[width=85mm]{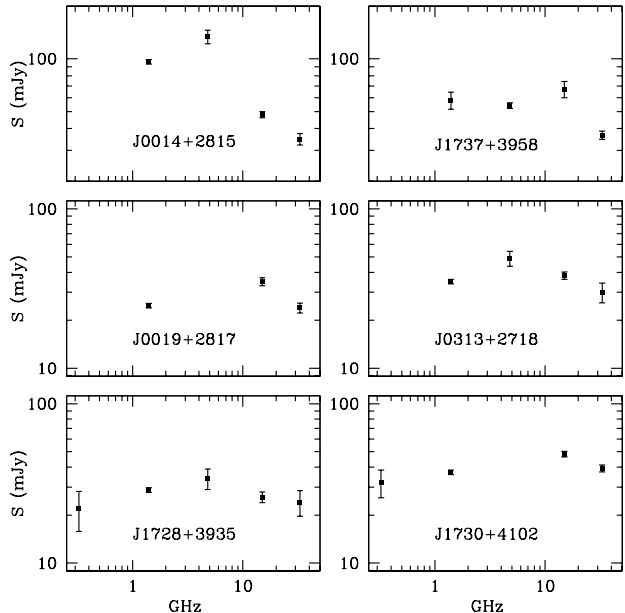}
\caption{New GPS source candidates found in the VSA sample.}
\label{f30}
\end{figure}

Finally, we attempt to identify sources with spectra peaking at a
frequency between 1.4 and 33\,GHz, by the requirements $\alpha_L>0$,
$\alpha_H<0$ and $\Delta\alpha<-0.3$: there are 10 sources satisfying
these conditions. Only 2 of them (0024+2911 and 1526+4201) are
included in the 8 GPS sources identified by C05. Six of the remaining
sources could be possible GPS candidates and their spectra are plotted in
Figure \ref{f30}. Further data are required to confirm their GPS shape.

\section{Sources with upturn spectrum in the 33--GHz VSA sample}

\begin{table*}
\caption{List of VSA sources with an upturn spectrum. The flux densities are
at the frequencies: 325\,MHz from the WENSS; 1.4\,GHz from NVSS; 4.8\,GHz from
GB6; 15\,GHz from RT; 33\,GHz from VSA. The reported spectral indices are
calculated using the VSA flux densities uncorrected by the selection bias. 
The last column reports the frequency of the minimum in spectra, as explained
in the text.}
\begin{tabular}{@{}lcccccccccccc@{}}
\hline
name & & \multicolumn{5}{c}{S(mJy)} & & $\alpha_L$ & $\alpha_H$ & 
$\alpha_H-\alpha_L$ & & $\nu_m$ \\
 & & 325\,MHz & 1.4\,GHz & 4.8\,GHz & 15\,GHz & 33\,GHz & & & & & & GHz \\
\hline
0013+2834 & & -- &  39.8$\pm$1.3 &  23.8$\pm$3.7 &  28.0$\pm$2. &
36.8$\pm$1.8 & & -0.42 & 0.27 & 0.69 & & 4.8 \\
0023+2928 & &  85.$\pm$7. &  41.8$\pm$1.3 &  26.2$\pm$5.2 &  20.3$\pm$2. &
35.5$\pm$1.9 & & -0.31 & 0.71 & 1.02 & & 15 \\
0024+2724 & & 318.$\pm$30. & 122.7$\pm$3.7 &  41.4$\pm$4.1 &  24.0$\pm$2. &
29.7$\pm$2.8 & & -0.73 & 0.27 & 1.00 & & 15 \\
0728+5325 & & 89.$\pm$7. & 60.$\pm$1.8 & 36.8$\pm$4.1 & 49.0$\pm$2. & 
52.9$\pm$2.7 & & -0.40 & 0.16 & 0.56 & & 4.8 \\
0728+5431 & &   72.$\pm$7. &  44.1$\pm$1.4 &  40.0$\pm$4.6 &  64.8$\pm$2. &
112.4$\pm$2.7 & & -0.08 & 0.65 & 0.73 & & 4.8 \\
0931+3049 & &  111.$\pm$8. &  67.6$\pm$2.1 &  23.3$\pm$3.1 &  24.0$\pm$2. &
32.6$\pm$6.7 & & -0.87 & 0.10 & 0.97 & & 4.8 \\
0944+3115 & &  111.$\pm$8. &  56.2$\pm$2.1 &  39.5$\pm$5.1 &  18.0$\pm$2. &
25.3$\pm$2.6 & & -0.46 & 0.43 & 0.89 & & 15 \\
0946+3050 & & 62.$\pm$7. & 37.8$\pm$1.2 & 20.4$\pm$3.3 & 28.9$\pm$2. &
28.5$\pm$2.8 & & -0.50 & 0.15 & 0.55 & & 4.8 \\
0946+3309 & & 791.$\pm$38. &  351.5$\pm$12.1 & 135.5$\pm$11.9 &  25.0$\pm$2. &
40.6$\pm$2.5 & & -1.09 & 0.62 & 1.71 & & 15 \\
0950+3201 & &  105.$\pm$7. &  56.2$\pm$2.1 & -- & 16.7$\pm$2. & 21.1$\pm$3.6
& & -0.51 & 0.30 & 0.81 & & 15 \\
1215+5154 & &   16.$\pm$6. &  16.4$\pm$0.6 & -- & 14.1$\pm$2. & 26.0$\pm$6.0
& & -0.06 & 0.78 & 0.84 & & 15 \\
1219+5408 & & 123.$\pm$8. & 42.6$\pm$1.3 & -- & 13.$\pm$2. & 29.$\pm$10. & &
-0.50 & 1.02 & 1.52 & & 15 \\
1221+5429 & & 45.$\pm$6. & 36.1$\pm$1.1 & 14.6$\pm$3.1 & 32.7$\pm$2. & 
20.7$\pm$2.5 & & -0.74 & 0.09 & 0.83 & & 4.8 \\
1229+5147 & &  579.$\pm$24. & 175.8$\pm$5.3 &  41.3$\pm$4.2 &  41.1$\pm$2. &
41.3$\pm$3.5 & & -1.18 & 0.00 & 1.18 & & 4.8 \\
1240+5441 & &  951.$\pm$39. & 223.0$\pm$6.7 &  59.9$\pm$6.5 &  20.0$\pm$2. &
27.8$\pm$6.2 & & -1.02 & 0.41 & 1.43 & & 15 \\
1533+4107 & & 53.$\pm$6. & 33.4$\pm$1.4 & 20.4$\pm$3.4 & 30.$\pm$2. & 
28.1$\pm$2.3 & & -0.40 & 0.12 & 0.52 & & 4.8 \\
1539+4217 & &  130.$\pm$8. &  49.0$\pm$1.5 &  25.3$\pm$3.5 &  31.0$\pm$2. &
34.0$\pm$1.6 & & -0.54 & 0.14 & 0.68 & & 4.8 \\
1723+4206 & &  682.$\pm$28. & 240.6$\pm$7.2 &  76.6$\pm$7.2 &  17.6$\pm$2. &
24.4$\pm$2.4 & & -1.07 & 0.41 & 1.48 & & 15 \\
1733+4034 & &  249.$\pm$12. & 84.1$\pm$3.0 & 38.8$\pm$4.9 &  16.0$\pm$2. &
20.8$\pm$4.4 & & -0.69 & 0.35 & 1.04 & & 15 \\
1734+3857 & &  478.$\pm$20. & 796.4$\pm$23.9 & 522.7$\pm$46.0 & 734.0$\pm$2. &
939.3$\pm$1.9 & & -0.34 & 0.31 & 0.65 & & 4.8 \\
\hline
\end{tabular}
\label{t1}
\end{table*}

\begin{figure*}
\includegraphics[width=165mm]{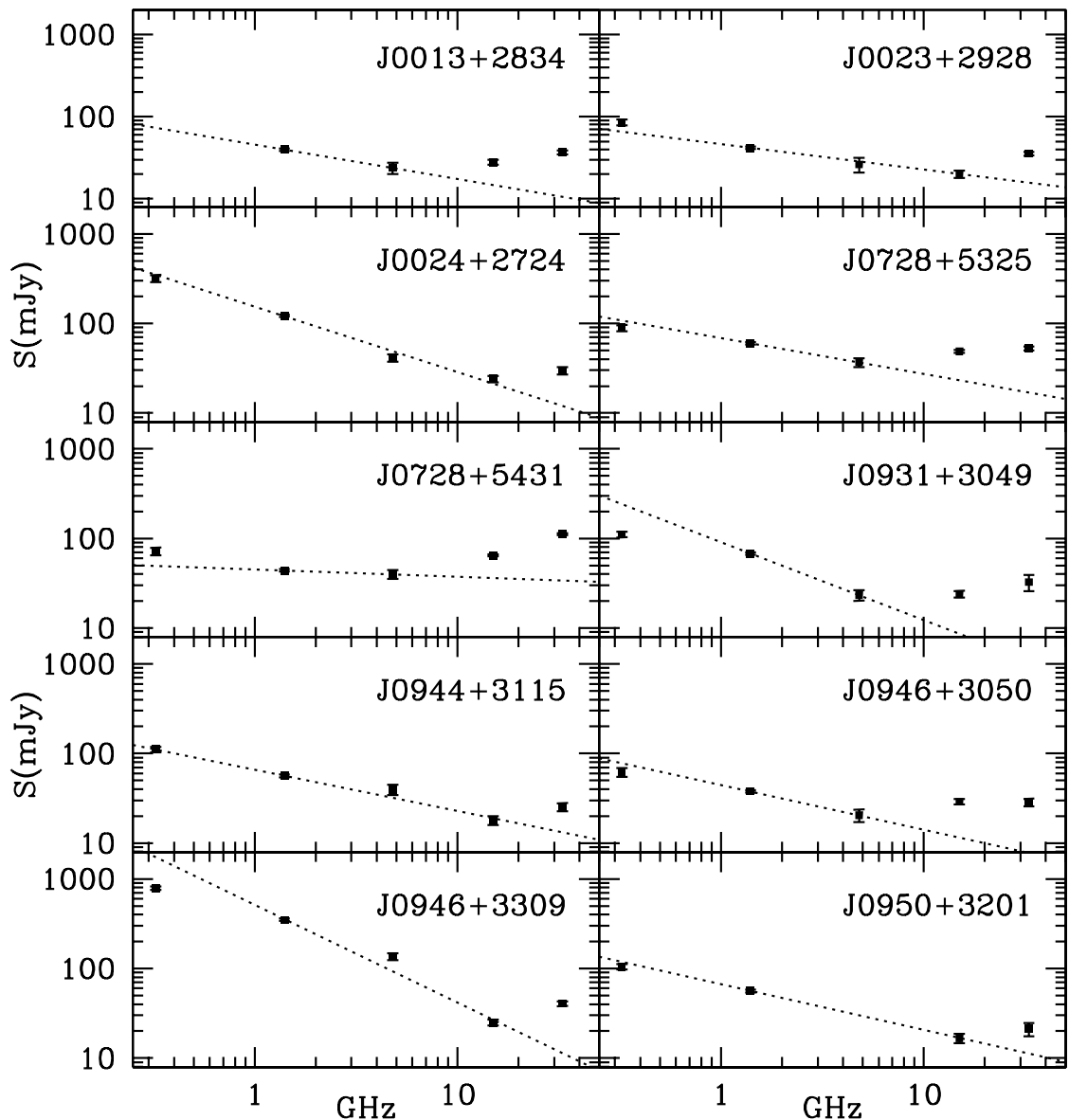}
\caption{Sources with upturn spectra. Dotted lines are the power laws that best
fit spectra at low frequencies.}
\label{f3a}
\end{figure*}
\begin{figure*}
\includegraphics[width=165mm]{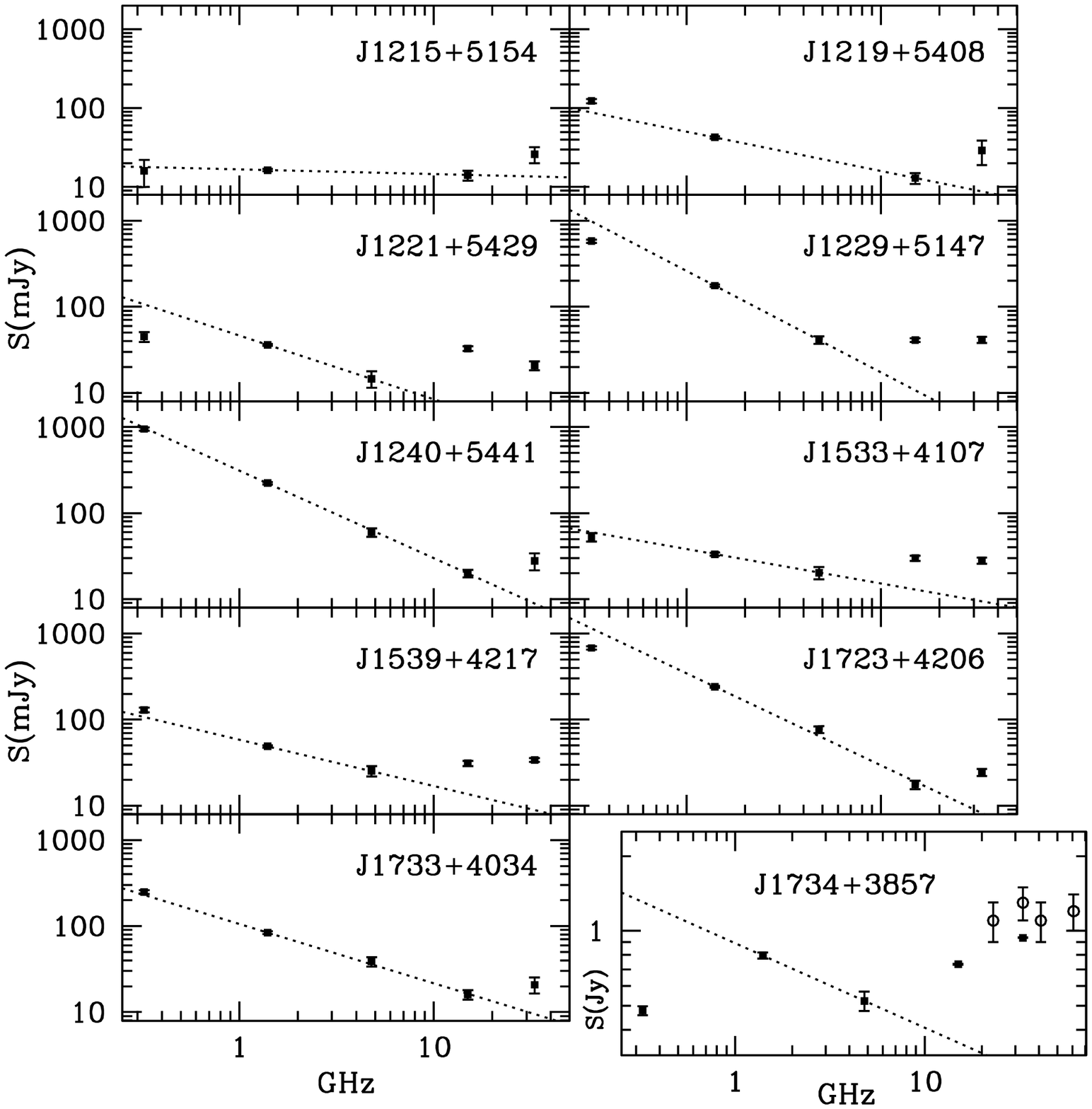}
\caption{Sources with upturn spectra (as in Figure \ref{f3a}). For source
1734+3857 we plot also the flux densities obtained by \citet{lop07} using the
3--year WMAP data (open points).}
\label{f3b}
\end{figure*}

Having previously characterised the source spectra, in this section we look for
objects with upturn spectra, or in other terms, with a minimum in the spectrum
at $\nu_{m}=4.8$ or 15 GHz. We say that a source has an upturn spectrum if
$\alpha_L<0$ and
$\alpha_H\ge0$, in agreement with \citet{sad06}, but with the additional
condition $\alpha_H-\alpha_L>0.5$, in order to minimize spurious cases. Over a
total of 105 sources, we find 27 objects with an upturn spectrum, corresponding
to 26\% of the sample (easily identified in Figure \ref{f3}). While in
\citet{sad06} most of the objects classified as upturn are flat--spectrum at
low frequencies with $\alpha_L\sim0$, in the VSA sample there are only 8 upturn
sources with $\alpha_L>-0.5$. This means that the majority of them are
steep--spectrum sources at low frequencies that become flat or inverted at
$\nu>5\,$GHz, with a strong variation in the spectral index (in some cases
$\ga1$).

The different survey resolutions (\,$<$1\,arcmin for
WENSS, NVSS and RT and $\sim$3\,arcmin for VSA and GB6) could affect
the shape of spectra in the case of extended sources. Among the
selected upturn sources, we find four that are resolved by NVSS, with the
major axis between 44 and 67\,arcsec.  Because the RT resolution is
25\,arcsec, the 15--GHz flux density could be underestimated, even if
integrated fluxes are considered. In particular, two extended sources,
0010+3044 and 1240+5334, are characterized by a peculiar spectrum with
a strong fall at 15\,GHz ($\alpha_{4.8}^{15}<-1.5$) that could be
explained if power on angular scales higher than 25\,arcsec is missed
by RT.  Some caution is therefore needed for these objects and they
are excluded from the analysis.

It is important to rule out that upturn
spectra could arise from bias in the 33--GHz flux density. It is known, in
fact, that for flux--limited surveys the flux density of faint sources (i.e.,
close to the flux limit of the survey) could be overestimated due to the
Eddington bias \citep{edd40}. Using Bayes' theorem, we can calculate the
maximum likelihood estimator of the flux density for a source with detected
flux density $S_0$ (\citealt{hog98}; \citealt{lop07}) as
\begin{equation}
S_{\rm ML}={S_0 \over 2}\Bigg[1+\sqrt{1-{4\beta \over r^2}}\Bigg]~~~~~~~~
{\rm with}~~r\ge2\sqrt{\beta}\,,
\label{e1a}
\end{equation}
where $r$ is signal--to--noise ratio of the detected flux density and $\beta$
is the slope for the differential source count distribution. For the VSA sample
the estimated slope is $\beta=2.34$ (C05), implying that the
Eq. \ref{e1a} can be applied only to sources with signal--to--noise higher than
3 (this condition is verified by $\sim97\%$ of the VSA sources). In the few
cases where S/N$\la3$, we consider $S_{\rm ML}=S_0/2$.

Now using the maximum--likelihood flux densities at 33\,GHz we calculate the
new values for $\alpha_L$ and $\alpha_H$ (the Eddington bias can be supposed
negligible in the other surveys because of the high S/N ratio, at least for the
VSA objects): after removing extended sources, 20 objects (19\% of the sample)
still verify the upturn conditions. We list them in Table \ref{t1}.

As discussed by C05, there is a deliberate deficiency of bright sources in the
VSA fields. The first fields to be observed were selected to avoid sources
predicted to be brighter than 250 mJy at 33 GHz; later fields used a more
relaxed limit of 500 mJy. As a consequence, the fields are statistically
representative of the whole sky only between 20 and 46 mJy for first fields and
between 20 and 114 mJy for the later fields.  Limiting to $S\le46\,$mJy we find
17 upturn sources over a total of 86, corresponding to 20\%, well in agreement
with the percentage from the all sample. A smaller percentage is obtained in
the area of 0.024\,steradian, where there are 7 upturn sources over 43 (16\%)
in the range 20--46\,mJy and 2 over 14 (14\%) between 46 and 114\,mJy.

In Figures \ref{f3a}--\ref{f3b} we show the spectra for the 20
upturn sources found in the VSA sample. For a visual inspection of spectra at
low frequencies we include the flux densities at 325\,MHz as measured by the
Westerbork Northern Sky Survey (WENSS, \citealt{ren97}). We plot also the
low--frequency best--fit power law: as expected for these sources, the
extrapolation of the low--frequency power law to high frequencies generally
underestimates the actual fluxes at 33\,GHz, in some cases up to a factor of 
10, like for the objects 0931+3049 and 1229+5147. 
It is remarkable how, in general, the flux density at
325\,MHz fits well the power law extrapolation from data at frequencies about
one order of magnitude higher, confirming the reliability of the power law
approximation for radio spectra at low frequencies. In few cases a flattening 
at 325\,MHz is observed, due to a possible effect of synchrotron
self--absorption or a low--energy cutoff in the electron energy spectrum
\citep{car91}.

An interesting case is source 1734+3857, plotted in Figure \ref{f3b}: it is the
brightest object in the VSA sample with a flux density a bit less than 1\,Jy.
This source has been detected in the WMAP 23, 33, 41 and 61--GHz
channels with a flux density around 1.1\,Jy (\citealt{hin06}; \citealt{lop07}),
confirming the rising spectrum at $\nu>5\,$GHz observed by the VSA measurement.

\subsection{Variability in upturn sources}

Since observations of the surveys were carried out at different
epochs, variability of radio sources could affect the shape of spectra. A
question is therefore if the observed upturn spectra could be fictitiously
generated by high variability in VSA sources. Two different cases can be
discussed according to the value of $\nu_m$.

For upturn sources with spectral minimum at $\nu_m=15\,$GHz, only the VSA flux
density does not agree with a spectral power--law fit. Because observations of
the RT and VSA are separated by no more than two years, such sources should be
high variable in a time scale of 1--2 years, and the VSA observations should
correspond to a peak in the flux density. VSA observations cover a time scale
from few months up to six months. As discussed by C05, the variability of 72
sources could be examined: the majority of them (about 70\%) varies by less
than 25 per cent of the mean flux density, a 20\% between 25 and 50 per cent
and 11\% more than 50 per cent. Only three upturn sources with $\nu_m=15\,$GHz
are found in these 72 objects, and they show to have varied by less than 20 per
cent. On the contrary, variability of at least 50 per cent is required to
produce changes in the spectral index, calculated between 15 and 33\,GHz,
bigger than 0.5, i.e. enough to explain upturn spectra. Moreover, high variable
objects are usually associated to flat--spectrum sources (see the Figure 9 of
C05 where a strong correlation between variability and spectral index is
found). This is confirmed by \citet{bol06}, who carried out a study of
variability in a time scale between 1.5 and 5 years: no steep--spectrum sources
have been classified as variable whilst 50 per cent of the flat--spectrum
sources have varied (up to a 70\% of the mean flux density). Also \citet{sad06}
found that at 20\,GHz the general level of variability is quite low in a
one--year interval, with only five sources varied by more than 30\%.
To conclude, it seems quite unlikely that variability could explain upturn
spectra with minimum at 15\,GHz.

On the other hand, the total set of data covers nearly twenty years (the GB6
survey was taken in 1986--87, WENSS in 1991--93 and NVSS between 1993 and 1996,
RT in 2000--2001 while VSA in 2002--03). Variability on time scales of years
could be relevant for sources with spectra upturning at $\nu_m=4.8\,$GHz:
because there are no
observations following these sources on time scales of ten years or more, we
are not able to exclude a significant long--scale variability in their flux
density. We consider the ``worst'' case, in which the flux densities measured
by GB6 at 4.8\,GHz (the most distant in time from VSA measurements) correspond
to a minimum in the source emission, giving a steeper low--frequency spectral
index. Ignoring the 4.8--GHz fluxes, we take $\alpha_L=\alpha_{1.4}^{15}$: over
the 10 upturn sources with the minimum at 4.8\,GHz only 2 still satisfied our
upturning conditions, even if 7 of them keep an upturn shape with the
less strict conditions $\alpha_L<0$, $\alpha_H>0$ and $\Delta\alpha\ga0.3$.
Therefore, in principle, a long--scale variability could explain the spectral
curvature observed in sources with minimum at $\nu_m=4.8\,$GHz (but not in ones
with $\nu_m=15\,$GHz) requiring a large fractional variability, in many
cases of the order of $100$ per cent at 4.8\,GHz. Because at the moment there
are no observational data supporting this hypothesis and because variability
is not able to explain all the upturn spectra found in the VSA sample,
we are still confident that the observed upturn spectra represent intrinsic
properties of radio sources. Finally, we observe that in most of upturn sources
the WENSS flux densities at 325\,MHz are in very well agreement with
extrapolations based on the estimated $\alpha_L$, indicating no variability on
a time scale of years at these frequencies.

\section{Discussion on flattening/upturn sources}

Although not explicitly considered, sources with flattening or upturn spectra
are consistent with the unified scheme for radio--loud AGNs (e.g.,
\citealt{orr82}, \citealt{bar89} and
\citealt{urr95}). According to this picture, spectral properties of the AGN
radio emission depend on the degree of the alignment of relativistic jets
with the line of sight (hereafter the ``viewing angle''): in a simplistic way,
steep spectra are observed in misaligned objects, where radio emission is
dominated by extended
optically--thin lobes; on the other hand, if the line of sight is enough close
to the AGN jet, the optically--thick and beamed emission of the AGN compact
core emerges with the typical flat spectrum \citep{pea85}. The critical viewing
angle $\theta_c$ that defines the division between flat-- and steep--spectrum
objects at GHz frequencies is usually considered around $10\degr$ for quasars
and $30\degr$ for BL\,Lacs (\citealt{ghi93}; \citealt{jac99}). Therefore, for
angles $\theta$ not much larger than $\theta_c$, we expect a transition from
steep to flat spectra at some frequency $\nu\ga5\,$GHz, when the compact
emission begins to dominate over the extended one (whose spectrum falls rapidly
with the frequency). \citet{jac99} employed this idea in order to extrapolate
the number counts of radio sources at 151\,MHz to 5\,GHz, introducing the
``beamed'' contribution of parent populations of blazars. This picture is
in agreement with high--resolution observations: AGN radio structures are
commonly resolved into three basic components (nuclear core, large scale jets
and outer lobes), while extended radio structures without a contribution from a
core are rare (\citealt{pea88}). Moreover, the relative importance of the AGN
core emissions is observed to increase at higher and higher frequencies (see,
e.g., \citealt{bac04} for Cygnus A and \citealt{kha07} for a sample of
Fanaroff-Riley\,II).

Now we discuss more quantitatively how the unified scheme applies to our case.
We consider the observed flux density of VSA sources to be the sum of two
components, an extended steep--spectrum and a flat--spectrum core--dominated
emission. For simplicity, we suppose that the VSA sample is composed of
quasars, being the dominant population at that frequency and flux range. Let
$R_c^{obs}$ be the ${\rm observed}$ core--to--extended flux ratio at 5\,GHz. At
a frequency $\nu>5\,$GHz the core component dominates the total emission if
\begin{equation}
R_c^{obs}=\Big({S_c \over S_{ext}}\Big)_{5\,{\tt GHz}}>
\Big({\nu \over 5\,{\tt GHz}}\Big)^{\alpha_{ext}-\alpha_c},
\label{e1}
\end{equation}
where $\alpha_{ext}$ and $\alpha_c$ are the spectral index of the two
components, extended and compact. In the following, we adopt as spectral 
indices the typical values for steep-- and flat--spectrum sources at low
frequencies, $\alpha_{ext}=-0.75$ and $\alpha_c=-0.1$ \citep{dez05}. The value
for $\alpha_{ext}$ could be conservative at frequencies $\nu>10\,$GHz because
it does not take into account possible steepening due to the electron ageing.

Objects with flattening or upturn spectrum in the VSA sample should be
characterized by $S_c/S_{ext}>1$
at 33\,GHz, corresponding to $R_c^{obs}>0.3$. A relevant fraction of them,
moreover, shows flattening/inverted spectra already at 15\,GHz, requiring
$R_c^{obs}>0.5$. If the core emission comes out from a relativistic jet of
bulk velocity $\beta c$, the observed flux density is Doppler--enhanced by
$\delta^{p-\alpha_c}$, where $\delta=[\gamma(1-\beta\cos\theta)]^{-1}$ is the
Doppler factor and $\gamma=(1-\beta^2)^{-1/2}$ is the Lorentz factor. The
parameter
$p$ depends on the emission model and is equal to 3 for a moving, isotropic
source and 2 for a continuous jet (see \citealt{ghi93} and \citealt{urr95} for
a detailed discussion). Then, the relation between the observed ($R_c^{obs}$)
and the intrinsic--unbeamed ($f$) core--to--lobe flux ratio for a source at
redshift $z$ is given, after the K--correction, by
\begin{equation}
R_c^{obs}=f\,\delta^{p-\alpha_c}\,(1+z)^{\alpha_c-\alpha_{ext}}.
\label{e2}
\end{equation}
In the unified scheme, the value of $f$ is the same for quasars and for their
parent population, the FR\,II galaxies. Because there is no evidence that the
parameter $f$ correlates with the radio power of FR\,II galaxies \citep{jac99},
we can use the best--fit FR\,II value $f=5\times10^{-3}$ \citep{urr95} for both
steep-- and flat--spectrum quasars, independently of their luminosity.

Using Eq. \ref{e2}, now we are able to provide an indicative estimate of
$R_c^{obs}$ for quasars observed at a viewing angle
$10\degr\le\theta\le20\degr$: supposing a Lorentz factor between 2 and 15, we
obtain
\begin{equation}
\begin{array}{cclc}
\theta=10\degr & \delta=[3.4,\,5.7] & R_c^{obs}=[0.3,\,1.5] & p=2 \nonumber \\
& & R_c^{obs}=[1.0,\,8.3] & p=3 \nonumber \\
\theta=15\degr & \delta=[1.8,\,3.7] & R_c^{obs}=[0.05,\,0.4] & 
p=2 \nonumber \\ 
& & R_c^{obs}=[0.09,\,1.5] & p=3 \\
\theta=20\degr & \delta=[1.1,\,2.7] & R_c^{obs}=[10^{-2},\,0.15] & 
p=2 \nonumber \\ 
& & R_c^{obs}=[10^{-2},\,0.4] & p=3
\end{array}
\end{equation}
The K--correction is of little relevance, being a factor $\sim1.6$ for $z=1$
(the redshift we use). Fixing the viewing angle and for the considered
parameters, the value of $R_c^{obs}$ scatters over a range of a factor 30. In
any case, we note that for $\theta=10\degr$ $R_c^{obs}$ is compatible with the
values required to have an upturn spectrum, independently of the $\gamma$ and
$p$ used. This is not the case when $\theta$ is near to $20\degr$ (which should
be considered as upper limit) and upturn spectra are possible only for low
values of the Lorentz factor ($\gamma\la5$) and $p=3$. The case of
$\theta=15\degr$ may be more conservative because the condition that
$R_c^{obs}=0.3$ is verified independently of $p$, at least for $\gamma\la10$.
Therefore, according to this model and for the range of frequencies
investigated, a viewing angle of $\theta\la15\degr$ is required to observe
quasars with a flattening spectrum.

\subsection{Predicted number of flattening/upturn sources in the VSA sample}

A prediction of how many upturn sources should be expected in a sample selected
at 33\,GHz is quite difficult, not only for the scattering in the physical
parameters used in Eq. \ref{e2}, but also for selection effects in
flux--limited samples. However, following the previous discussion, a
qualitative estimate is possible. We want to stress that nearly all the VSA
upturn sources (18/20) have $\alpha_L<-0.3$. Sources with
$-0.5<\alpha_L<-0.3$, although not strictly steep spectrum according the usual
definition, can be considered a transition between lobe-- and core--dominated
quasars. In the following we do not make a distinction between them.

We suppose that all the sources with steep spectra at low frequencies observed
at viewing angles $10\degr\le\theta\le15\degr$ show a flattening or rising
spectrum at $\nu\ga5$\,GHz. If lobe--dominated quasars are in general observed
at $10\degr\le\theta\le40\degr$ \citep{ghi93}, we should find that about 1/6 of
steep--spectrum sources have a flattening or upturn spectrum at high
frequencies (in 
the hypothesis that the probability to observe a radio quasars is independent
of $\theta$). This estimate should be only applied to samples selected at few
GHz. In flux--limited samples selected at high frequencies, this percentage
could be higher because of missing very steep spectrum sources. As example, in
the VSA sample we have found 18 upturn objects out of a total of 69 sources
which are steep ($\alpha_L<-0.3$) at low frequencies.

In order to establish if the number of flattening/upturn sources expected from
the model is in agrement with the number observed in the VSA sample, we proceed
in the following way:
\begin{itemize}

\item we suppose that for flattening/upturn sources the average
``high--frequency'' spectral index is $\langle\alpha_H\rangle=0$.

\item Using the NVSS and GB6 surveys we select inside the VSA fields all
sources with $\alpha_{1.4}^{4.8}<-0.3$ and flux density higher than 20\,mJy
at 4.8\,GHz. We find 370 objects.

\item Extrapolating the flux density to 15\,GHz using $\alpha_{1.4}^{4.8}$, we
find that 149 of these 370 sources have $S_{15}\ge20\,$mJy. According to the
above discussion, we expect that 1/6 of them (i.e. 25
objects) have flat or inverted spectrum at high frequencies ($\alpha_H\sim0$),
and are observed in the VSA sample.

\item Moreover, the rest of the 221 steep--spectrum sources (with
$S_{4.8}\ge20\,$mJy but the extrapolated $S_{15}<20\,$mJy) can
also be observed in the VSA sample if their spectra are flat starting from
$\nu=4.8\,$GHz. These objects correspond to the previous cases with
$\nu_m=4.8\,$GHz, and they are more or less one half of the total number of
flattening/upturn sources. Therefore, considering a fraction of 1/12, we have a
further contribution of 18 objects.

\end{itemize}

To resume, the number of sources with flattening/upturn spectrum predicted by
our estimates in the VSA area at 33\,GHz is 43. This value should be compared
with the number of steep--spectrum ($\alpha_L<-0.3$) sources with
$\Delta\alpha=\alpha_H-\alpha_L\ge0.3$ found in the VSA sample. This difference
is in fact approximately the change of the spectral index in the case
$S_c>S_{ext}$ at 33\,GHz (using the $\alpha_c$ and $\alpha_{ext}$ assumed
above). As seen in Table \ref{t0}, there are 43 sources with these
characteristics, as from our predictions. A smaller number, 37,
is found if we correct the 33--GHz VSA flux densities by the Eddington bias.

If we want to limit our predictions only to sources with an upturn behaviour,
i.e. rising at high frequencies, we should require that $\alpha_c>0$ and
$S_c\gg S_{ext}$ at 33\,GHz. The fraction of flat--spectrum sources with
$\alpha>0$ can be obtained from samples of sources whose emission is dominated
by the AGN compact core: \citet{ric06} showed that for a complete sample of
flat--spectrum sources at
5\,GHz, the distribution of the spectral index calculated between 2.7 and
5\,GHz is well described by a Gaussian with $\langle\alpha\rangle=-0.1$ and
$\sigma_{\alpha}=0.3$. These values are similar to parameters describing the
spectral index distribution at low frequencies for flat--spectrum sources in
the VSA sample (see Figure \ref{f300}). Using these parameters in a Gaussian
distribution, we find that about 38 per cent of spectral indices are higher
than 0. This means that 16 of the 43 predicted objects with flattening spectrum
is expected to be upturn between 5 and 33\,GHz, again in agreement with the
results from the VSA sample.

\section{Upturn sources from high--frequency surveys}

In this section we search for sources with upturn spectra in other surveys at
frequencies comparable or higher than the VSA one. The fraction of ``upturn''
sources is expected to depend on the selection frequency and the flux-density
range of surveys \citep{kel68}. A direct comparison with the VSA results should
take into account this.

\begin{itemize}

\item {\bf WMAP source catalogue}.
The WMAP mission has produced the first all--sky surveys of extragalactic
sources at 23, 33, 41, 61 and 94\,GHz and at Jy flux densities
(\citealt{ben03}; \citealt{hin06}). In this interval of frequencies, spectral
properties of extragalactic sources are still poorly investigated.
From 3--years WMAP maps, \citet{hin06} provided a catalogue of 323 objects
above a flux limit of about 1\,Jy. This catalogue was improved by \citet{lop07}
using a non--blind method based on the second member of the Mexican Hat Wavelet
Family. They found 368 sources with a 5--$\sigma$ detection in at least one
WMAP channel and with a completeness flux limit of around 1.1\,Jy.

We complement the WMAP flux densities by measurements from low--frequency
surveys: in the northern hemisphere from the NVSS and GB6 surveys; in the
southern hemisphere from the SUMSS \citep{mau03} and PMN (\citealt{wri94};
\citealt{wri96}) surveys at 843\,MHz and 4.85\,GHz respectively. We use these
data to define the low--frequency spectral index ($\alpha_L$) for WMAP sources.
Objects without a counterpart both in NVSS/SUMSS and in GB6/PMN or without a
clear identification have been discarded from the analysis: 248 objects remain,
114 with low--frequency spectral index $\alpha_L<0$ and only 22 with
$\alpha_L\le-0.5$. If the high--frequency spectral index is calculated between
4.8\,GHz and the lowest frequency at which a source has been detected by WMAP,
we find 25 upturn sources (5 with $\alpha_L\le-0.5$) according to the previous
definition. Some examples of these with counterparts in the multi--frequency
catalogue by \citet{kue81} or in the ATCA observations (\citealt{ric06};
\citealt{mas07}) are shown in Figure \ref{f5}.

The detection of objects with upturn spectrum in the WMAP catalogue is an
important confirmation of the presence of this class of sources also at Jy flux
levels. Their fraction is about 10\%, lower than in the VSA sample. This is not
unexpected due to the different range of flux densities considered by the two
experiments. At Jy level, in fact, the source population is dominated by
flat--spectrum sources \citep{dez05} and the fraction of objects with inverted
spectrum at low frequency is particularly high. In the WMAP catalogue more than
$50\%$ of sources have $\alpha_L>0$, compared to $17\%$ in the VSA sample.

For a better comparison between the two samples we consider only the WMAP
sources with a 5--$\sigma$ detection in the 33--GHz channel, in order to 
achieve a
sample selected at the same frequency of the VSA. Then, we compute $\alpha_L$
and $\alpha_H$ as in section 3.1 (using the 23--GHz WMAP channel instead of the
15--GHz frequency). Now, if we limit the comparison only to objects with
decreasing spectrum at low frequency ($\alpha_L<0$), we find about $20\%$ of
upturn sources in the WMAP catalogue, very close to $23\%$ in the VSA sample.

\begin{figure}
\includegraphics[width=85mm]{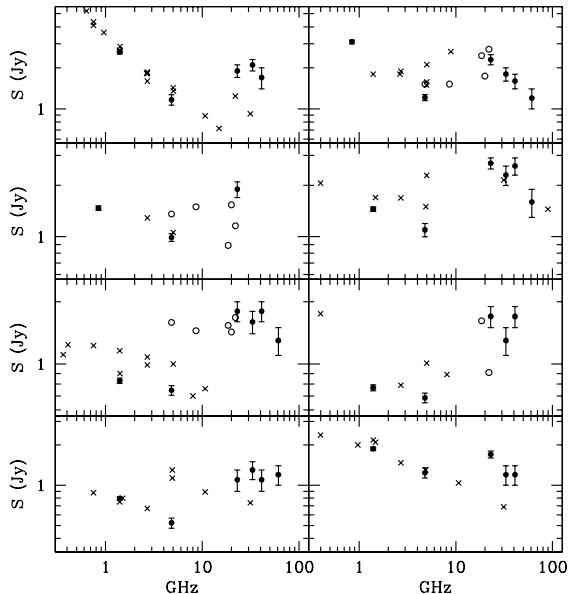}
\caption{Examples of upturn spectra in the WMAP catalogue. Counterparts in the
\citet{kue81} and in the ATCA samples are shown as cross points and open
squares respectively. Full points are from WMAP and from SUMSS/NVSS and PMN/GB6
at low frequencies.}
\label{f5}
\end{figure}

\item {\bf The 20\,GHz ATCA Survey}. A pilot survey of the Australia Telescope
20\,GHz survey was presented by \citet{sad06}, consisting of a flux--limited
sample ($S\ge100\,$mJy) of 173 radio sources selected at 20\,GHz, including
near--simultaneous flux measurements at 4.8, 8.6 and 18\,GHz. In the paper,
using also the flux densities provided by the SUMSS catalogue \citep{mau03},
they carried out an accurate analysis of source spectra in the frequency range
0.8--20\,GHz: about 20\% of source spectra were classified upturn (defined as
$\alpha_{0.8}^{5}<0$ and $\alpha_{8}^{18}>0$). However, introducing our 
stricter condition that $\alpha_{8}^{18}-\alpha_{0.8}^{5}\ge0.5$, the fraction
is reduced to 14\%.

Recently, a complete sample of bright sources ($S_{20GHz}>0.5\,$Jy), carried
out with ATCA, was presented by \citet{mas07}. For 218 of them, near
simultaneous observations at 4.8, 8.6 and 20\,GHz were also provided.
\citet{mas07} studied the source spectral behaviour in the frequency range
5--20\,GHz: a general steepening in spectra is observed, while very few upturn
sources are found (only three according our definition; see their Figure 4).
Extending the analysis to the NVSS and SUMSS measurements at 1.4 and 0.8\,GHz,
the number of sources with upturn spectra increases to 19 (8\% of the sample),
indicating that their spectral minimum is at $\la5\,$GHz. However, as for the
WMAP case, if we exclude objects with inverted spectrum at low frequency (that
are about 60\% of the sample), the fraction of upturn sources is of the order
of 20\%, well in agreement with the VSA and WMAP one.

\item {\bf Radio follow--up of the 9C survey}. \citet{bol04} carried out a
follow--up of 176 sources from the 15--GHz 9th Cambridge survey in the
frequency range [1.4,\,43]\,GHz. In particular, they made simultaneous
measurements at 1.4, 4.8, 22 and 43\,GHz with the VLA and at 15\,GHz with the
Ryle Telescope.
As discussed by the authors (see also \citealt{wal07}), there is a clear trend
for spectra to be steeper at the high frequencies: an average steepening is
already observed at 22\,GHz, but it becomes especially strong in the range
15--43\,GHz (the median spectral index is --0.89), with a large number of
objects with extremely steep spectra, less than $-1.5$. We find only 22 (14\%)
inverted sources between 4.8 and 22\,GHz and very few objects inverted up to
43\,GHz or upturn.

This strong steepening at $\nu>15\,$GHz has not been observed in other
surveys (ATCA measurements found a steepening in 20--95\,GHz spectral indices,
however the median values is --0.39; see \citealt{sad07}) and it is clearly in
disagreement with VSA results. Such a discrepancy may be related to the
different resolution of the instruments, e.g. a few arcsec for the VLA at
22--43\,GHz, 25\,arcsec for the Ryle Telescope and just 3\,arcmin for the VSA
source subtractor.

\end{itemize}

\section{Discussion and conclusions}

An accurate control of the contribution of extragalactic radio sources
at microwave wavelengths will be extremely important for future CMB
experiments at arcminute resolution like Planck and for experiments
aimed to the detection of the cosmological B--mode polarization
\citep{tuc05}. The VSA source--subtractor observations provide a very useful
nearly--complete catalogue of ``faint'' (20--100\,mJy) extragalactic
radio sources at a frequency used for CMB measurements. In this paper, using
additional measurements at lower frequencies, we investigate the spectral
properties of the VSA catalogue in the frequency range 1.4--33\,GHz.

The most interesting result we find is related to the high--frequency
spectral behaviour of sources whose spectrum is steep at GHz frequencies.
Typical models for this source population (see, e.g., the recent work by
\citealt{dez05}) predict a spectral steepening at high frequencies due to the
rapid ageing of high--energy electrons, responsible of the emission at
$\nu\gg1\,$GHz. On the contrary, from the 33--GHz VSA catalogue we observe
that the spectrum of the majority of them flattens or even becomes inverted at
$\nu\ga5\,$GHz. In particular, in about 50 per cent of steep-spectrum sources
the spectral flattening at high frequencies is more than 0.5. Moreover, we find
objects with upturn spectra, i.e. steep at low frequencies but rising between
$5<\nu<33\,$GHz: 20 sources clearly show such spectral behaviour, corresponding
to 19 per cent of the total sample.

The large fraction of sources characterized by spectral flattening and the lack
of correlation between flux density and spectral curvature are indications that
we are observing intrinsic properties of radio sources, and not the result of
selection effects or bias. Variability can affect the shape of spectra observed
in measurements that are not simultaneous. However, it is quite unlikely that
variability is responsible of the average spectral flattening found in VSA
steep--spectrum sources. In fact, it would require large variations in flux
densities on time scales of both 1--2 years and a decade, for a source
population usually observed to exhibit little variability (see, e.g.,
\citealt{sad06}, \citealt{bol06}).

In the paper, we show that source spectra with flattening or upturn behaviour
are completely consistent with the ``unified model'' 
for AGNs: steep spectra in radio sources can be physically understood as
emission from extended optically--thin radio lobes, while a flat--spectrum
component arises from the central compact and optically--thick core of radio
sources. Due to the rapid decrease in intensity of the diffuse emission with
frequency, the emission from the compact core of AGNs may start to dominate
at frequency $\nu\gg1\,$GHz if the radio jet axis lies enough close to the line
of sight \citep{jac99}. According to the typical physical parameters for
quasars, it can occur for $\nu\ga5$\,GHz when the viewing angle is
$\theta\la15\degr$.

Finally, VSA data confirm that a simple power--law extrapolation of
low--frequency flux densities to mm wavelengths is not reliable: this is not
only because a fraction of sources predicted at high frequencies are not found,
but also because a significant number of sources that would not be expected are
detected. For example, using a simple extrapolation of the spectral slope
between 1.4 and 4.8\,GHz to 33\,GHz, we find that 34 sources present in the VSA
catalogue would be predicted to be fainter than the flux limit of 20\,mJy and
not included in the sample.
From surveys in literature, the percentage of these ``unexpected'' objects is
observed to steadily increase with the frequency. While it is roughly 10\% at
15\,GHz (see \citealt{tay01}) and 18\% at 20\,GHz (see \citealt{sad06}), this
percentage increases up to 32\% in the VSA sample.

We conclude that radio sources with flattening or upturn spectrum could be an
important contaminant to high--resolution CMB observations providing to an
extra contribution to source number counts respect to predictions based on
surveys at 1--5\,GHz. At 33\,GHz this extra contribution is not
negligible: limited to the 20 upturn sources found in the VSA sample, we
substitute their measured flux density with the one obtained from a
low--frequency extrapolation; in this case, the VSA number counts at 33\,GHz
would keep the same slope as found by C05 but rescaled down by 15 per cent.
This is for flux densities $20\le S\la100\,$mJy. At higher fluxes the fraction
of upturn sources is observed to be lower, around 10 per cent in the WMAP
and ATCA source catalogues at Jy levels. If confirmed,
it would suggest an extra contribute from upturn sources of about 10--15\% to
temperature fluctuations generated by extragalactic radio sources at 33\,GHz.
At higher frequencies the
relevance of this class of sources is expected to increase.

In the near future a number of new ground--based surveys at 20--30\,GHz will be
available, as, for example, the super--extended version of VSA, the Australia
Telescope 20\,GHz survey and the One Centimetre Receiver Array (OCRA)
project. They will be able to more deeply investigate radio source spectra in
large intervals of fluxes and to confirm the results discussed in this paper.
However, observations at frequencies higher than 30\,GHz are required for a
better characterization of upturn sources. In this sense, the ESA's Planck
satellite will be extremely important, carrying out all--sky surveys from 30 to
860\,GHz with a sensitivity several times better than WMAP, and extending
high--frequency radio surveys to fluxes of few hundred of mJy.

\vskip 0.7truecm \noindent {\it Acknowledgements}. MT thanks to Luigi
Toffolatti for the careful reading of the manuscript and helpful comments and 
Elizabeth Waldram for the useful discussions.

\label{lastpage}

\end{document}